\documentstyle{aipproc}
\begin{document}

\title{Commutation Relations in Mesoscopic Electric Circuits}
\author{You-Quan Li}
\address{
Institut f\"ur Physik, Universit\"at Augsburg, D-86135 Augsburg, Germany\\
and Department of  Physics, Zhejiang University, Hangzhou 310027, China\\
email: yqli@physik.uni-augsburg.de}

%\lefthead{LEFT head}
%\righthead{RIGHT head}
\maketitle   

\begin{abstract}
In the talk, I briefly demonstrate the quantum theory for mesoscopic
electric circuits and its applications.
In the theory, the importance of the charge discreteness
in a mesoscopic electric circuit is addressed. 
As a result, a new kind of commutation relation  
for electric charge and current occurred inevitably.
The charge representation, canonical current representation and 
pseudo-current  representation are discussed extensively. 
It not only provides a concrete realization of
mathematical models which discuss the space quantization in high energy 
physics and quantum gravity but also presents a sequence of 
applications in condensed matter physics from a different point of view.
A possible generalization to coupled circuits is also proposed.
\end{abstract}

\section*{Introduction}

The dramatic achievement in nanotechnology has aroused tremendous 
developments in experimental physics in mesoscopic scale. 
Miniaturization of integrated circuits is undoubtedly a 
persistent trend for electronic device community. 
A theory for mesoscopic circuits was proposed by Li and Chen,
in which the charge discreteness is first introduced in the quantization 
of electric circuits\cite{LiC}. The possibility of space-time discreteness  
was early considered by Snyder\cite{Snyder} who indicated that the Lorentz
invariance do not exclude quantized (discrete) space-time, 
it was also argued
by Li\cite{Li95} from the finiteness of the observed universe. 
S. Mantecinos, I. Saavedra, and O. Kunstmann \cite{MSaavedra} 
discussed the commutation relations of 
\cite{Snyder} arguing that it may be related to physics in high energy scale
($10^9-10^{12}$eV). Numerous attempts to the argument of exiting  minimal 
position uncertainty were made \cite{Garay} on the basis of various 
considerations in string theory as well as in quantum gravity. Actually,
the approach of \cite{LiC} not only provides a concrete realization of
mathematical models for exploring the space quantization
in high energy physics and quantum gravity but also presents a sequence of 
applications in condensed matter physics from a different point of view. 
For example, the persistent current is solved by regarding
the mesoscopic metal ring as the circuit of a pure L-design.
Application of the theory to a pure C-design gives rise to 
the Coulomb blockade solution\cite{Li97}. 

\section*{Basic definitions and commutation relations}

Let $\hat{q}$ denote for the {\it charge} operator, and $\hat{p}$ 
for the canonical conjugation of the charge satisfying
$\bigl[ \hat{q},\hat{p} \bigr]=i\hbar$.
We call $\hat{p}$ the {\it canonical current} operator 
since it is not only the canonical conjugation of charge but also  
the current operator in the quantization approach\cite{Louisell}
where the charge was considered as a continuous variable.
Taking into account of the discreteness of electronic charge 
in quantization procedure,
we must impose that
the eigenvalues of the self-adjoint operator $\hat{q}$ take discrete
values\cite{LiC}, i.e. $\hat{q}|n\rangle=nq_e|n\rangle$
where $n \in {\sf \, Z \!\!\! Z \, } $ (set of integers) and
$q_e  = 1.602 \times 10^{-19}$ coulomb, the elementary electric charge.  
We therefore introduce a minimum `shift operator'
${\displaystyle
\hat{Q}:=e^{iq_e\hat{p}/\hbar}}$ in charge space, 
which satisfies\cite{LiC}
\begin{equation}
\bigl[ \hat{q},\hat{Q} \bigr]=-q_e\hat{Q},\;\;\hat{Q}^{-1}=\hat{Q}^\dagger.
\label{eq:qQ}
\end{equation}
These relations determine the structure of the whole Fock space,
accordingly, 
$\hat{Q}^{+}| n\rangle= e^{i\alpha_{n+1}}|n+1\rangle$,
$\hat{Q}|n\rangle=e^{-i\alpha_{n}}|n-1\rangle$
where $\alpha_n $'s being undetermined phases. 
The Fock space for the algebra (\ref{eq:qQ}) differs from the well known 
Fock space for the Heisenberg-Weyl algebra 
because the spectrum of the former is isomorphic to the set of
integers ${\sf\,Z\!\!\!Z\,}$ but that of the later is isomorphic to 
the set of non-negative integers ${\sf\,Z\!\!\!Z\,}^{+}+\{0\}$.
Since $\{|n\rangle|n\in{\sf\,Z\!\!\!Z\,}\}$ spans a Hilbert space
and $\hat{q}$ is self-adjoint,
both the completeness $\sum_{n\in{\sf\,Z\!\!\!Z\,}}|n\rangle\langle n|=1$
and the orthogonality $\langle n|m\rangle=\delta_{n m}$ faithful.

The {\it quasi-current} $\hat{J}$ for a mesoscopic circuit is defined by 
$\hat{J}=-i\hbar(\hat{Q}^{1/2}-\hat{Q}^{-1/2})/q_e$ 
which reduces to the canonical current in the limit $q_e\rightarrow 0$. 
The Hamiltonian of a mesoscopic LC-circuit is given by\cite{LiC}
\begin{equation}
\hat{H}=-\frac{\hbar^2}{2L}\hat{J}^2
 +\frac{1}{2C}\hat{q}^2+\varepsilon\hat{q},
\label{eq:Hamiltonian}
\end{equation}  
where $\varepsilon$ stands for the voltage source, 
$L$ for inductance, and $C$ for capacity of the circuit.
Using eq.(\ref{eq:qQ}) we easily obtain the new commutation relations for
the quasi-current operator,
\begin{equation}
\bigl[ \hat{q}, \hat{J} \bigr]=i\frac{\hbar}{2}\hat{K}, \;\;
\bigl[ \hat{q}, \hat{K} \bigr]=-i\frac{q_e^2}{2\hbar}\hat{J},
\label{eq:JK}
\end{equation} 
where an auxiliary operator $\hat{K}=\hat{Q}^{1/2}+\hat{Q}^{-1/2}$ 
is introduced. Obviously eq.(\ref{eq:JK}) obeys the SU(2) algebra after
rescaling the operators.
In terms of $\hat{K}$ and $\hat{J}$ we can define a
useful operator $\hat{P}=\hat{J}\hat{K}^{-1}$ which we
call the {\it pseudo-current} operator. Obviously the
pseudo-current also reduces to canonical current in the limit
$q_e\rightarrow 0$. With the help of (\ref{eq:JK}), we obtain
the following commutation relations,
\begin{equation}
\bigl[\hat{q},\hat{P}\bigr]=
 i\hbar\Bigl(1+(\frac{q_e}{2\hbar})^2\hat{P}^2\Bigr).
\label{eq:qP}
\end{equation}
Similar kind of commutation relation was considered earlier in \cite{Snyder}
in searching the possibility of space-time discreteness. 
From the commutation relation (\ref{eq:qP}) one will have a uncertainty
relation \cite{MSaavedra,Kempf} for the charge and pseudo-current\cite{LiC},
which is different from the conventional Heisenberg uncertainty relation.

The definition of {\it physical current} $\hat{I}$ arises from the
Heisenberg equation 
$\hat{I}=d\hat{q}/dt=(1/i\hbar)\bigl[ \hat{q}, \hat{H} \bigr]$.
For the $LC$-design circuit, one can immediately obtain \cite{Flores},
\begin{equation}
\hat{I}=-i\frac{\hbar}{2q_eL}(\hat{Q}-\hat{Q}^\dagger).
\label{eq:current}
\end{equation} 

\section*{Pseudo-current representation}

We consider the pseudo-current representation
$\hat{P}|\eta\rangle=\eta|\eta\rangle$. 
The differential realization of commutation
relation (\ref{eq:qP}) is given by \cite{Snyder}
\begin{equation}
\hat{P}=\eta, \;\; 
\hat{q}=i\hbar\Bigl(
  1+(\frac{q_e}{2\hbar}\eta)^2\Bigr)
   \frac{\partial}{\partial\eta}.
\label{eq:eta}
\end{equation}
Obviously, $\int d\eta\psi^*(\eta)\hat{q}\phi(\eta)$ fails in guaranteeing
the charge operator being self-adjoint. The factor
$(1+(\eta q_e/2\hbar)^2)^{-1}$ in the measure on the pseudo-current
space is therefore required \cite{Kempf} to cancel the 
corresponding factor of $\hat{q}$
in this representation. The inner product must be so defined,
\begin{equation}
\langle\psi|\phi\rangle
 =\int^{\infty}_{-\infty}\frac{d\eta}{1+(\frac{q_e}{2\hbar}\eta)^2}
  \psi^*(\eta)\phi(\eta)
\label{eq:inner}
\end{equation}
that both $\hat{q}$ and $\hat{P}$ could be self-adjoint.

The completeness is given by 
\begin{equation}
\int^{\infty}_{-\infty}\frac{d\eta}{1+(\frac{q_e}{2\hbar}\eta)^2}
 |\eta\rangle\langle\eta|=1.
\label{eq:complete}
\end{equation}
Consequently, the inner product of two eigenstates of the pseudo-current
operator yields 
\begin{equation}
\langle\eta'|\eta\rangle
 =\Bigl(1+(\frac{q_e}{2\hbar}\eta)^2\Bigr)\delta(\eta-\eta').
\label{eq:innerEta}
\end{equation}

The eigen-equation of charge operator in this representation reads
\begin{equation}
i\hbar\Bigl(1+(\frac{q_e}{2\hbar}\eta)^2\Bigr)
 \frac{\partial}{\partial\eta}\psi_q(\eta)
  =q\psi_q(\eta),
\label{eq:eigenEq}
\end{equation}
where $\psi_q(\eta):=\langle\eta|\psi_q\rangle$. The deferential equation 
(\ref{eq:eigenEq}) is solved by 
\begin{equation}
\psi_q(\eta)=(\frac{q_e}{2\pi\hbar})^{1/2}\exp
 \bigl(-iq\frac{2}{q_e}\tan^{-1}(\frac{q_e}{2\hbar}\eta)\bigr),
\label{eq:q-eigen}
\end{equation}
which has been normalized. It is interesting to evaluate their inner product
\begin{equation}
\langle\psi_q|\psi_{q'}\rangle=\frac{q_e}{(q'-q)\pi}\sin(\frac{q'-q}{q_e}\pi).
\label{eq:q-inner}
\end{equation}
This clearly brings about a orthogonal catastrophe because the eigenstate
of a self-adjoint operator with different eigenvalues must be mutually
orthogonal. Actually, it can be avoided provided that 
$q'-q=nq_e$. We conclude that the electric charge must be quantized 
(other eigenvalues are not physically permitted). 

A natural choice is $q=nq_e$, then the transformation from charge 
representation to pseudo-current representation is easily derived,
\begin{eqnarray}
\langle\eta|\psi\rangle&=&\sum_{n=-\infty}^{\infty}
     \langle\eta|n\rangle\langle n|\psi\rangle \nonumber\\
      \,   &=&(\frac{q_e}{2\pi\hbar})^{1/2}\sum_{n=-\infty}^{\infty}
              \langle n|\psi\rangle e^{-in\Theta(\eta)},
\label{eq:qP-transf}
\end{eqnarray}
where $\Theta(x)=2\tan^{-1}(xq_e/2\hbar)$.
Multiplying (\ref{eq:qP-transf}) by 
$e^{in'\Theta(\eta)}/[1+(\eta q_e/2\hbar)^2]$
and integrating with respect to $\eta$ give rise to the inverse
transformation:
\begin{equation}
\langle n|\psi\rangle=(\frac{q_e}{2\pi\hbar})^{1/2}
 \int_{-\infty}^{\infty}\frac{d\eta}{1+(\frac{q_e}{2\hbar}\eta)^2}
  e^{in\Theta(\eta)}\langle\eta|\psi\rangle.
\label{eq:Pq-transf}
\end{equation}

\section*{Canonical current representation}

In the canonical current space $\hat{p}|p\rangle=p|p\rangle$, 
the $\hat{q}$ and $\hat{p}$ are realized
by $\hat{p}=p$, $\hat{q}=i\hbar\partial/\partial p$. 
The eigen-equation of charge operator is,
\[
i\hbar\frac{\partial}{\partial p}\psi_q(p)=q\psi_q(p),
\]
which is solved by plane waves
$\psi_q(p)=e^{-iqp/\hbar}$.
Obviously, the periodic condition in p-space,
$\psi_q(p+2\pi\hbar/q_e)=\psi_q(p)$
should be imposed so that the charge is quantized (discrete)
$q/q_e=n$, consequently,
\begin{equation}
\psi_n(p)=\langle p|n\rangle=e^{-inpq_e/\hbar}
\label{eq:q-wf}
\end{equation}

The transformation from charge representation to canonical current 
representation is easily obtained, 
\begin{eqnarray}
\langle p|\psi\rangle&=&
 \sum_{n=-\infty}^{\infty}\langle p|n\rangle\langle n|\psi\rangle\nonumber\\
  \,  &=&\sum_{n=-\infty}^{\infty} e^{-inpq_e/\hbar}\langle n|\psi\rangle.
\label{eq:qp-transf}
\end{eqnarray}
Multiplying eq.(\ref{eq:qp-transf}) with $e^{in'pq_e/\hbar}$
and integrating with respect to $p$, we get the inverse transformation,
canonical current representation to charge representation:
\begin{equation}
\langle n|\psi\rangle=\frac{q_e}{2\pi\hbar}\int_{-\pi\hbar/q_e}^{\pi\hbar/q_e}
 dp\langle p|\psi\rangle e^{inpq_e/\hbar}.
\label{eq:p-n}
\end{equation}
In the p-space, the Hamiltonian for a mesoscopic LC-design circuit becomes
\begin{equation}
\hat{H}=-\frac{\hbar^2}{q_e^2L}\left[\cos(\frac{q_e}{\hbar}p)-1\right]
  -\frac{\hbar^2}{2C}\bigl(\frac{\partial}{\partial p}
   +i\frac{C}{\hbar}\varepsilon\bigr)^2 -\frac{C}{2}\varepsilon^2.
\label{eq:Hamiltonian-p}
\end{equation}
The advantage of the canonical current representation is that
the Schr\"odinger equation for the Hamiltonian (\ref{eq:Hamiltonian-p})
becomes the standard Mathieu equation after a unitary transformation.
The wave function was solved in terms of periodic Mathieu functions, 
and the energy spectrum was expressed by the eigenvalues of
Mathieu equation. The details can be find in \cite{LiC}.

\section*{Applications}

In Coulomb blockade experiments,
the mesoscopic capacity may be relatively very small 
(about $10^{-8}F$) but the inductance of a macroscopic circuit 
connecting to a source is relatively large
because it is proportional to the area which the circuit spans. 
We can neglect the term 
reversely proportional to $L$ in (\ref{eq:Hamiltonian}), 
and study the equation for a pure C-design. Because a mesoscopic metal
ring can be regarded as a pure L-design, we can also study the persistent 
current on a mesoscopic ring.

\subsection*{Coulomb blockade}

The Schr\"odinger equation for a pure C-design reads
\begin{equation}
\Bigl(\frac{1}{2C}\hat{q}^2 -\varepsilon\hat{q}
  \Bigr)|\psi\rangle=E|\psi\rangle,
\end{equation}
where $\varepsilon$ is an adiabatic voltage source. 
The Hamiltonian and charge operator commute each other,
so $|n\rangle$ is the eigenstate with energy   
$ E=(n q_e - C\varepsilon)^2/2C 
  -C\varepsilon^2/2$,
where both the charge quantum number and the voltage source
are involved.
The relation between charge $q$
and the voltage $\varepsilon$ for the {\it ground} state is given by
\begin{equation}
q=\sum_{m=0}^{\infty}
   \left\{ \theta[\varepsilon -(m+\frac{1}{2})\frac{q_e}{C}]
            -\theta[-\varepsilon -(m+\frac{1}{2})\frac{q_e}{C}]
             \right\}q_e
\label{eq:chargestep}
\end{equation}
where $\theta(x)$ is the step function. 
The corresponding eigenstate is 
\begin{equation}
|\psi(\varepsilon)\rangle_{ground}=
   \sum_{m=-\infty}^{\infty}
     \left\{\theta[\varepsilon -(m-\frac{1}{2})\frac{q_e}{C}]
            -\theta[\varepsilon -(m+\frac{1}{2})\frac{q_e}{C}]
             \right\}|m\rangle.
\end{equation}
The dependence of the current on time is obtained by taking 
derivative 
\begin{equation}
\frac{dq}{dt}=\sum_{m=0}^{\infty}q_e
   \left\{\delta[\varepsilon -(m+\frac{1}{2})\frac{q_e}{C}]
            +\delta[\varepsilon + (m+\frac{1}{2})\frac{q_e}{C}]
             \right\}\frac{d\varepsilon}{dt}.
\end{equation}
Clearly, the current is of a form of sharp pulses which occurs
periodically (with periodicity $q_e/C$)
according to the changes of voltage. 

\subsection*{Persistent current}

The Schr\"{o}dinger equation for a pure L-design in the presence of 
magnetic flux is given by,
\begin{equation}
-\frac{\hbar^2}{2q_e^2L}(e^{-i\frac{q_e}{\hbar}\phi}\hat{Q}
  +e^{i\frac{q_e}{\hbar}\phi}\hat{Q}^+ -2)
    |\psi\rangle=E|\psi\rangle.
\label{eq:gh}
\end{equation}
It is obtained on the basis of gauge covariance \cite{LiC}.
The eigenstates can be simultaneous eigenstates of $\hat{p}$,
eq.(\ref{eq:gh}) is solved by the eigenstate 
$|p\rangle=\sum_{n\in{\sf \,Z\!\!\!Z\,}}
\kappa_n e^{i n q_e p/\hbar}|n\rangle$
($\kappa_n :=\exp(i\sum_{j=1}^n \alpha_j)$)
with the energy spectrum: 
\begin{equation}
E(p,\phi) = \frac{2\hbar^2}{q^2_eL} \sin^2
\Bigl( \frac{q_e}{2\hbar}(p-\phi)
\Bigr).
\end{equation}
It oscillates with respect to
$\phi$ or $p$. Differing from the usual classical pure L-design, the
energy of a mesoscopic quantum pure L-design can not be large than
$2\hbar^2/q_e^2L$.
Clearly, the lowest energy states are those states with
$ p = \phi + n h/q_e $,
the eigenvalues of the electric current $\hat{I}$
of ground state can be obtained \cite{LiC}.
The electric current on a mesoscopic  circuit of pure L-design
is not null in the presence of a magnetic flux (except
$\phi = nh/q_e$).
This is a quantum characteristic property.
The persistent current in a mesoscopic L-design
is an observable quantity periodically depending on the flux $\phi$.
In terms of the inductance of mesoscopic
metal ring, $L=8\pi r (\frac{1}{2}\ln\frac{8r}{a} - 1)$
where $r$ is the radius of the ring and $a$ is the radius of the metal wire,
the formula for persistent current on a mesoscopic ring is obtained
\begin{equation}
I(\phi) = \frac{\hbar}{8\pi r 
   (\frac{1}{2}\ln\frac{8r}{a} - 1)q_e}
     \sin(\frac{q_e}{\hbar}\phi).
\end{equation}
Differing from the conventional formulation of the persistent
current on the basis of quantum dynamics for electrons, This formulation
presented a  method from a new point of view. Formally, the $I(\phi)$
is a sine function with periodicity of
$\phi_0 =h/q_e$. 
But either the model that the electrons move freely in an ideal 
ring \cite{Cheung}, or the model that the electrons have hard-core
interactions between them \cite{LiMa2} can only give the sawtooth-type
periodicity. Obviously, the sawtooth-type function is only
the limit case for $q_e /\hbar \rightarrow 0 $.

\section*{Coupled circuits}

The above discussions are based on a single mesoscopic circuit. Let
us consider the case that several circuits coupled to each other by
mutual inductances, associated capacities or any other kind of coupling.
Because of quantum tunneling and quantum fluctuations, the charges
in individual circuit is no longer precisely measurable. The charge
on each circuit are not good quantum numbers, and therefore the 
charge operators $\hat{q_j}$ are expected to be noncommutative.
A natural generalization of the formulation for single circuit is easily
carried out in the pseudo-current representation, namely,
 \begin{equation}
\hat{P}_j=\eta_j, \;\;\;
\hat{q_j}=i\hbar\Bigl(1+(\frac{q_e}{2\hbar})^2\vec{\eta}\,^2\Bigr)
 \frac{\partial}{\partial\eta_j}.
\label{eq:multi}
\end{equation}
It is easily to obtain the following commutation relation,
\begin{equation}
\bigl[ \hat{q}_i, \hat{q}_j  \bigr]=i\frac{q_e^2}{2\hbar}M_{ij},
\label{eq:q-ij}
\end{equation}
where $M_{ij}:=\hat{P}_i\hat{q}_j-\hat{P}_j\hat{q_i}$.
The charge operators for individual circuit are generally noncommutative
for the non-vanishing  $\langle M_{ij}\rangle$.
This provides a concrete physical example of noncommutative geometry. 
Further studies are in progress.

\section*{acknowledgment}
  
The work is supported by AvH Stiftung, NSFC-19975040 and EYFC98.
I would like to thank J.C. Flores for communication and reference.  
Interesting discussions with O.W. Greenburg, M.S. Plyushchay, A. Solomon,
E.C.G. Sudarshan, A. Zee {\it et.al.} are also acknowledged.

\end{document}